\documentstyle[aps,preprint,epsf]{revtex}

\def\z2{{Z\!\!\!\!Z^2}}

\def\chib{{\overline\chi}}
\def\tri{{\bigtriangleup}}

\begin{document}
\setcounter{page}{0}
\def\footnoterule{\kern-3pt \hrule width\hsize \kern3pt}
\title{A LARGE $N$ CHIRAL TRANSITION ON A PLAQUETTE\thanks
{This work is supported in part by funds provided by the U.S.
Department of Energy (D.O.E.) under cooperative 
research agreement \#DF-FC02-94ER40818.}}

\author{Shailesh Chandrasekharan
\footnote{Email address: {\tt chandras@ctpa03.mit.edu}}}

\address{Center for Theoretical Physics \\
Laboratory for Nuclear Science \\
and Department of Physics \\
Massachusetts Institute of Technology \\
Cambridge, Massachusetts 02139 \\
{~}}

\date{MIT-CTP-2530, August 1996}
\maketitle

\thispagestyle{empty}

\begin{abstract}

We construct a model of a chiral transition using the well known
large N transition in two dimensional $U(N)$ lattice gauge theory.
Restricting the model to a single plaquette, we introduce
Grassmann variables on the corners of the plaquette with 
the natural phase factors of staggered fermions and couple them
to the $U(N)$ link variables. The classical theory has a continuous 
chiral symmetry which is broken at strong couplings, but is restored
for weak couplings in the $N\rightarrow \infty$ limit.


\end{abstract}

\vspace*{\fill}


\section{Introduction}

  Chiral transitions can occur in (non-Abelian) gauge theories coupled to 
fermions. These transitions can be continuous or discontinuous depending
on the specific dynamics of the model. If the up and down quark masses were 
exactly zero, we expect the real world to undergo such a transition at a 
finite temperature. It is further believed that the order of chiral 
transitions crucially depends on the number of quark 
flavors\cite{Pisarski}. If the strange quark was sufficiently light, the 
phase transition in the real world could be first order. These possibilities 
have created a lot of interest in studying chiral transitions in lattice
simulations\cite{Lattice}.

Our theoretical understanding of such transitions is essentially based on 
universality arguments\cite{Rajagopal}, which have worked quite well for many 
condensed matter systems. The present evidence from lattice simulations 
supports the predictions of universality. However, recently there have been 
some concerns\cite{Kogut} regarding the validity of such arguments for 
fermionic systems. Further, even if universality arguments are valid, 
there exist dynamics specific to the gauge fields that can
be important near the transition and cannot be determined in the
universal model. For example, it is well known that anomalous axial 
symmetry breaking effects can play an important role in determining the 
order of the transition\cite{Pisarski}. Recent lattice simulations
\cite{Chandras} have suggested a rapid evolution in the anomalous symmetry
breaking effects close to the chiral transition. Thus it is useful to 
understand the dynamics of the gauge fields that are responsible for chiral 
symmetry restoration. As a step in this direction it can be useful
to study simplified models, that have some non-trivial gauge dynamics.

  Chiral symmetry breaking in gauge theories is closely connected
to the small eigenvalue density of the Dirac operator. Recently some
progress has been made in understanding the spectrum of small eigenvalues of 
the Dirac operator in QCD\cite{Verbaar}. The models that have been studied 
are the so called random matrix models. Such models are motivated by the
chiral symmetries of the Dirac operator. It is possible that these models 
incorporate some dynamics of gauge fields at a crude level. When such models 
are extended\cite{Jackson} to include finite temperature effects, one
obtains a chiral transition. An interesting feature of the random matrix 
models is that the dynamical variable is typically an $N\times N$ matrix. The 
transitions are obtained in the thermodynamic limit which involves taking 
$N$ to infinity. In this sense, these can be regarded as large $N$ 
transitions of a matrix model. 

A question that immediately strikes
is whether there exist other known large N transitions that can be
interpreted as a chiral transition with an appropriate extension.
In this article we show that one can construct a chiral transition 
based on the large N transition discovered 
by Gross and Witten\cite{Gross} and Wadia\cite{Wadia} independently. 
They showed that the two dimensional $U(N)$ lattice gauge theory 
with the Wilson gauge action, has a large $N$ transition essentially due
to the non-analytic behavior of a single plaquette integral. 
Consequently the pure $U(N)$ lattice gauge theory with the Wilson gauge
action has a third order phase transition separating strong and weak
couplings. In the present work we show that, if we consider only a single 
plaquette, it is possible to include fermionic variables on the corners of 
the plaquette and turn the transition into a chiral transition. 
In the next two sections we define the theory more 
precisely and solve it in the large N limit, showing the existence
of the chiral transition. In the final section we make some observations
on the results obtained.

\section{The One Plaquette Theory}

  The theory we are considering is the well known one plaquette model,
which consists of four $N\times N$ unitary matrices living on the sides of 
a square(plaquette). To this we add Grassmann N-vectors, $\chi_n$,$\chib_n$ 
associated to the corners $n=1,2,3,4$. We will refer to them as fermionic 
variables as opposed to the unitary matrices that will be referred to as 
gauge(or link) variables. The unitary matrix that connects the adjacent 
corners $n$and $m$ will be represented as $U^{nm}$. The theory is described 
by the action
\begin{equation}
\label{eq:plaqact}
S[U,\chi,\overline\chi]\;=\; -{1\over g^2}Tr(U_p+U_p^\dagger) - 
\chib_n (D_{nm}+m \delta_{nm}) \chi_m,
\end{equation}
where $U_p\;=\;U^{12}U^{23}{U^{43}}^\dagger{U^{14}}^\dagger$. Note that
the sign in front of the gauge action is natural from lattice QCD. The
operator $D$ couples the fermionic variables on adjacent cites. We will
refer to $D$ as the Dirac operator in the present model. The non-zero 
components of the operator $D_{nm}$ are given by 
$D_{12}\;=\;\eta_{12}U^{12}$, $D_{23}\;=\;\eta_{23}U^{23}$,
$D_{34}\;=\;\eta_{43}{U^{43}}^\dagger$,
$D_{41}\;=\;\eta_{14}{U^{14}}^\dagger$, with the anti-hermitian  property
that $D_{mn}\;=\;-(D_{nm})^\dagger$. Thus in the matrix notation we have
\begin{equation}
D\;=\;\left( \begin{array}{cccc} 
	0 & \eta_{12}U^{12} & 0 & -\eta_{14}U^{14} \\
  	-\eta_{12}{U^{12}}^\dagger & 0 & \eta_{23}U^{23} & 0 \\
  	0 & -\eta_{23}{U^{23}}^\dagger & 0 & \eta_{43}{U^{43}}^\dagger \\
  	\eta_{14}{U^{14}}^\dagger & 0 & -\eta_{43}{U^{43}}^\dagger & 0  \\
\end{array}\right).
\end{equation}
The factors $\eta_{12},\eta_{23},\eta_{43}$ and $\eta_{14}$ could be
thought of as extra couplings in the theory. In order to fix these couplings
we use the analogy with staggered fermions. Such factors occur
naturally in the staggered fermion formulation\cite{Staggered} as phase
factors which are remnants of gamma matrices. In two dimension one encounters
phase factors denoted by $\eta_\mu(x)$, where $\mu=1,2$ represents the 
direction and $x=(x_1,x_2)$ the lattice site. For example one can chose 
$\eta_1(x)=1$ and $\eta_2=(-1)^{x_1}$. The well known staggered fermion 
action is then given by
\begin{equation}
\label{eq:stagg}
S_{staggered} = \sum_x \chib(x)\eta_\mu(x)\left(U_{x,x+\mu}\chi(x+\mu)
-U^\dagger_{x-\mu,x}\chi(x-\mu)\right).
\end{equation}
In the present model since we have a plaquette, we can think of it as an 
object in two dimension. Restricting the staggered fermion action 
eq.(\ref{eq:stagg}) to a single plaquette the factors 
$\eta_{12},\eta_{23},\eta_{43}$ and $\eta{14}$ can be naturally associated
with $\eta_\mu(x)$. Using this connection we obtain one choice of the 
factors to be
\begin{equation}
\label{eq:phases}
\eta_{12}=\eta_{43}=\eta_{14}=-\eta_{23}=1.
\end{equation}
We will see that with the definition of the plaquette action given in
eq.(\ref{eq:plaqact}), the above choice of the $\eta$ 
factors\footnote{Other equivalent choices would lead to the same results.} 
is necessary for the existence of the chiral transition.

We can now define the partition function of the theory by the usual integral 
over the gauge and fermionic variables
\begin{equation}
Z\;=\;\int [dU][d\chi][d\chib]\;\exp(-S[U,\chi,\chib]).
\end{equation}
It is obvious by construction that the theory is gauge invariant in the
the usual sense. Further the model also has a chiral symmetry, given by
\begin{equation}
\chi_n \rightarrow \exp(-i\theta (-1)^n)\chi_n,\;\;\;
\chib_n \rightarrow \chib_n\exp(-i\theta (-1)^n)
\end{equation}
in the chiral limit $m\rightarrow 0$. This is the remnant of the chiral 
symmetry usually present in a staggered fermion formulation. This suggests 
that the chiral order parameter, $\langle\chib\chi\rangle$, defined by
\begin{equation}
\label{eq:chcond}
\langle\chib\chi\rangle\;=\;{1\over Z}
\int [dU][d\chi][d\chib]\;\sum_n 
{1\over N}\chib_n\chi_n\;\exp(-S[U,\chi,\chib]),
\end{equation}
would vanish in the chiral limit. This will certainly be true for finite
$N$. However, as we will show below in the large N limit, the theory breaks 
chiral symmetry at strong couplings. Further, the well known large N 
transition of the pure gauge theory restores the chiral symmetry in the 
weak coupling.

\section{Large N Solution}
\label{sec3}

  We will follow the work Gross and Witten\cite{Gross} 
closely\footnote{with minor differences in notation} and calculate the chiral 
condensate defined in eq.(\ref{eq:chcond})
in the large N limit. It is important to recognize that 
the large N limit is not an approximation here, but is necessary for the 
transition. It is obvious that the above partition function will not show non-analytic properties when
N is finite. Further as usual the chiral limit must be considered only after 
the large N limit is taken. Hence we are naturally lead to study the large N 
solution to the theory. 

  Since the theory is gauge invariant we can go to a convenient gauge before
doing the integrals. We will fix the gauge such that $U^{23}=U^{14}=U^{43}=1$.
This naturally imposes the constraint that $U^{12}=U_p$. Using the properties
of the Haar measure we obtain,
\begin{mathletters}
\begin{eqnarray}
\label{eq:chcond1a}
\langle\chib\chi\rangle\;&=&\;{1\over Z}
\int [dU_p]\;{1\over N}Tr\left({1\over D+m}\right)
\exp\left({1\over g^2}Tr(U_p+U_p^\dagger)+Tr\;\log\;[D+m]\right),
\\
\label{eq:chcond1b}
Z\;&=&\;
\int [dU_p]\;\exp\left({1\over g^2}Tr(U_p+U_p^\dagger)+Tr\;\log\;[D+m]\right).
\end{eqnarray}
\end{mathletters}
The matrix $D$ is now given by
\begin{equation}
\label{eq:gfixD}
D\;=\;\left( \begin{array}{cccc} 
	0 & U_p & 0 & -1 \\
  	-U_p^\dagger & 0 & -1 & 0 \\
  	0 & 1 & 0 & 1 \\
  	1 & 0 & -1 & 0  \\
\end{array}\right),
\end{equation}
where it is important to remember that each element in the above $4\times 4$
matrix is an $N\times N$ matrix. The integrand in eq.(\ref{eq:chcond1a})
and eq.(\ref{eq:chcond1b}) only depends on the eigenvalues, 
$\alpha_i, i=1,...,N$ of $U_p$. This is obviously true for the part of the
integrand that depends only on the trace of $U_p$. To see that this is also
true for the remaining terms that depend on the matrix $D$, note 
that only eigenvalues of $D$ enter the integrand. It is easy
to compute the eigenvalues of $D$. Firstly note that the only 
$N\times N$ matrix in $D$ defined in eq.(\ref{eq:gfixD}), that is not a 
multiple of unity is $U_p$. Thus using the matrix $T\delta_{n,m}$, where $T$ 
is the $N\times N$ matrix such that 
$(T^\dagger U_pT)_{ij}\;=\;\delta_{ij} e^{i\alpha_i}$; 
$\alpha_i\in(-\pi,\pi]$, it is possible to
block diagonalize $D$ into $4\times 4$ matrices, $D^i$, each of which 
is associated with one eigenvalue $\alpha_i$ and is given by
\begin{equation}
\label{eq:blockD}
D^i\;=\;\left( \begin{array}{cccc} 
	0 & \exp(i\alpha_i) & 0 & -1 \\
  	-\exp(-i\alpha_i) & 0 & -1 & 0 \\
  	0 & 1 & 0 & 1 \\
  	1 & 0 & -1 & 0  \\
\end{array}\right).
\end{equation}
Thus the eigenvalues of D turn out to be
\begin{equation}
\label{eq:eigen}
\lambda^+_j = \pm i\;\sqrt{2(1+ \sin{\alpha_j\over 2})},\;\;\;
\lambda^-_j = \pm i\;\sqrt{2(1- \sin{\alpha_j\over 2})},\;\;j=1,2,...,N.
\end{equation}
Note that, due to the anti-hermitian nature of $D$ and the chiral symmetry 
discussed above, the eigenvalues are imaginary and come in pairs 
of opposite sign. Clearly if $\alpha_i=\pi$ $\lambda^-_i = 0$ and
$D^i$ has zero eigenvalues. This connection between $\alpha=\pi$ and zero
eigenvalues of $D$ plays an important role in the existence of the chiral 
transition being discussed.
The Haar measure simplifies to 
$[dU_p]\;=\;\hbox{const}\;dT\prod_i d\alpha_i\;\tri^2(\alpha_i)$, where
\begin{equation}
\tri^2(\alpha_i) \;=\;\prod_{i<j}\;\sin^2
\left|{\alpha_i-\alpha_j\over 2}\right|.
\end{equation}
Hence the integral over $T$ and various constants drop from expectation values
such as the chiral condensate. Further in the large $N$ limit, the integral 
is dominated by a single $U_p$ with a specific eigenvalue distribution. This 
distribution can be obtained using the saddle point condition,
\begin{equation}
\label{eq:saddle}
{2\over g^2} \sin\alpha_j
+ {4\sin\alpha_j\over m^4+4m^2+4\cos^2({\alpha_j\over 2})} \;=\;
\sum_{k\neq j}\cot\left|{\alpha_j-\alpha_k\over 2}\right|.
\end{equation}
Except for the second term on the left hand side, this is exactly the same
equation as in \cite{Gross}. The extra term comes from the extra fermionic 
determinant in the present model.  To get nontrivial results we take the 
$N\rightarrow \infty$ limit while keeping $G=N g^2$ fixed.
In this limit the extra fermionic piece plays no role
since it is suppressed by a power of $N$, a familiar effect 
related to the usual suppression of fermions in the large $N_c$ limit. Thus
in the large $N$ limit the saddle point equation, eq.(\ref{eq:saddle}), 
reduces to a continuum version that was studied in \cite{Gross}, namely
\begin{equation}
{2\over G} \sin\alpha(x)\;=\;
P\int_0^1\;dy\;\cot\left|{\alpha(x)-\alpha(y)\over 2}\right|,
\end{equation}
where $\alpha(x)$ is a nondecreasing function in the range $0\leq x\leq 1$
with $\alpha_i\;=\;\alpha(i/N)$. This equation can be solved by introducing
the eigenvalue density $\rho_{U_p}(\alpha)\;=\;dx/d\alpha$ which is 
nonzero in the interval $|\alpha| \leq \alpha_c, \alpha_c\leq \pi$. We have 
introduced the subscript $U_p$ to distinguish this eigenvalue density
from the one to be introduced shortly, namely the eigenvalue density of 
the operator $D$. It is the latter that plays an interesting role in the
calculation of the chiral condensate. Borrowing the results from 
\cite{Gross} we have
\begin{mathletters}
\label{eq:rhoup}
\begin{eqnarray}
\rho_{U_p}(\alpha)\;&=&\;{1\over 2\pi}[1+{2\over G}\cos\alpha],\;\;
|\alpha|\leq\alpha_c, \;\;\; G\geq 2, \\ 
\rho_{U_p}(\alpha)\;&=&\;{2\over G\pi}\cos(\alpha/2)
[\sin^2(\alpha_c/2)-\sin^2(\alpha/2)]^{1\over 2},\;\;
|\alpha|\leq\alpha_c,\;\;\; G\leq 2,\;\; 
\end{eqnarray}
\end{mathletters}
where $\alpha_c$ is equal to $\pi$ when $G\geq 2$ and is given by
$G/2 = \sin^2(\alpha_c/2)$ for $G\leq 2$. 

We can use these results now to 
calculate the chiral condensate. In the large $N$ limit one just has to evaluate 
${1\over N}Tr[1/(D+m)]$ in the background of $U_p$ whose eigenvalue distribution
is given by eq.(\ref{eq:rhoup}). We obtain
\begin{equation}
\label{eq:chicond2}
\langle\chib\chi\rangle\;=\;
\int_{-\alpha_c}^{\alpha_c}d\alpha \;\;\rho_{U_p}(\alpha)\left( 
{2 m\over {\lambda^+(\alpha)}^2+m^2} + {2 m\over {\lambda^-(\alpha)}^2+m^2}
\right),
\end{equation}
where the  values of $D$, namely $\lambda^+(\alpha)$ and
$\lambda^-(\alpha)$ are given by eq.(\ref{eq:eigen}) by just dropping the
subscript $j$. 

The reason for the chiral transition is now clear. Since
we need a non-zero density of small eigenvalues of $D$, we need to understand
the values of $\alpha$ for which one obtains small eigenvalues of $D$.
Note that $\lambda^+(\alpha)$ is always non-zero, while
$\lambda^-(\alpha)$ can become zero when $\alpha=\pi$. Thus for the chiral
condensate to be non-zero in the chiral limit, we need values of $\alpha$
close to $\alpha=\pi$ to contribute. For strong coupling, i.e. $G> 2$
this is clearly possible. However for weak coupling, i.e. $G\leq 2$,
one sees that $\rho_{U_p}(\alpha)=0$ for $\alpha= \pi$, which then suggests
the possibility of the vanishing of chiral condensate.

We define the eigenvalue density for the operator $D$ to be 
$\rho(\lambda)$. It is easy to show that 
$\rho(\lambda) = \rho(\alpha)(|d\alpha/d\lambda^+| + |d\alpha/d\lambda^-|)$,
where the first term is evaluated at $|\lambda^+|=\lambda$ and the second 
term is evaluated at $|\lambda^-|=\lambda$
\begin{equation}
\label{rhol}
\rho(\lambda)\;=\;{4\over \pi}\;\left\{ \begin{array}{cc}
\left(1 + {2\over G}[1-2(1-\lambda^2/2)^2]\right)/
\sqrt{4-\lambda^2} & \;\;\;0\leq\lambda\leq 2,\;\;\; G \geq 2, \\ \\
{2\lambda\over G}\sqrt{\left(G/2-(1-\lambda^2/2)^2\right)}
& \;\;\lambda_1\leq\lambda\leq \lambda_2,\;\;\;G \leq 2, \\ \\
0 & \hbox{otherwise},
\end{array}
\right.
\end{equation}
where $\lambda_1=\sqrt{2(1-\sqrt{G/2})}$ and 
$\lambda_2=\sqrt{2(1+\sqrt{G/2})}$. In figure 1 we plot $\rho(\lambda)$
for a few values of $G$ to illustrate the situation. Using the definition of 
$\rho(\lambda)$ we can write the chiral condensate to be
\begin{equation}
\label{eq:chicond3}
\langle\chib\chi\rangle\;=\;
\;\int_{0}^{2}d\lambda \;\;\rho(\lambda)\left( 
{2\;m\over \lambda^2+m^2} \right).
\end{equation}
In the chiral limit we obtain $\langle\chib\chi\rangle\;=\;\pi\;\rho(0)$.
We see that for $G\geq 2, \langle\chib\chi\rangle\sim (G-2)$, and for 
$G\leq 2, \langle\chib\chi\rangle=0$. Further, when $G=2$, we obtain
$\langle\chib\chi\rangle\sim m$. If we naively read off the critical 
exponents we find $\beta=1$ and $\delta=1$, quite different from mean field.
However, since we do not know of any universality arguments applicable to 
the model, we need not take the exponents too seriously. Further the
underlying transition is a third order transition\cite{Gross}.

\section{Conclusions}

 Some time ago it was shown by Eguchi and Kawai \cite{Eguchi} that in the 
large $N$ limit, $U(N)$ lattice gauge theory in d dimensions can be reduced to 
a model on a d dimensional hypercube. If this could be done in the
presence of staggered fermions one would obtain a model on the d dimensional
hypercube. In the present work we have formulated a simple model 
on the plaquette by hand. Thus it would be interesting to know if similar 
results can be obtained from the full 2-d theory in the spirit of the 
Eguchi-Kawai model where the final theory lives on the plaquette.
More over we know, the underlying phase transition, responsible for the
chiral transition just studied, is an artifact of the Wilson action used
for the plaquette variable. The phase transition can be eliminated if 
one uses the heat kernel action\cite{Menotti}. This limits the applicability
of the results obtained here. However, some qualitative features of a chiral
transition can be motivated using this example.

On a crude level, the results obtained here seem similar to the results of 
\cite{Jackson}. First, the transition studied here is driven by the gauge 
dynamics. The fermion determinant does not play any role; remember that in
the large $N$ limit the fermion determinant dropped out of the saddle point
equation eq.(\ref{eq:saddle}). In other words, the chiral transition occurs 
even in the quenched theory. Next, the density of eigenvalues of the Dirac 
operator develops a gap for small eigenvalues above the 
transition (see figure 1). One might wonder if these are generic features of 
chiral transitions involving gauge fields. In QCD, it is well known that the 
number of flavors plays an important role in determining the order of the 
transition. For example, it is believed that for one flavor there is no chiral 
transition\cite{Pisarski}, due to the presence of the anomaly. This suggests 
that there will not be any transition in the quenched case, since the absence 
of the fermion determinant is likely to enhance the small eigenvalues of the
Dirac operator and enhance the effects of chiral symmetry breaking.
If this is true, we can expect the chiral transition to be driven mainly by 
the fermion determinant. It is through the fermion determinant that the 
number of flavors enters the problem. Thus the present model and the random 
matrix models do not reflect these expected features of the QCD chiral phase 
transition.

  Another interesting observation is the role of the staggered fermion phases
defined in eq.(\ref{eq:phases}). Notice that in the weak coupling phase the 
gap in $\rho(\lambda)$ for small $\lambda$ is related to the gap in the 
eigenvalue density $\rho_{U_p}(\alpha)$ at $\alpha=\pi$. This can be traced 
back to the zero eigenvalues of $D^i$, defined in eq.(\ref{eq:blockD}),
which exist when $\alpha_i=\pi$ as discussed in section~\ref{sec3}.
If we had chosen all the signs in eq.(\ref{eq:phases}) to 
be positive, the zero eigenvalues of $D^i$ would be related to $\alpha_i=0$.
Since this would not effect the saddle point equation,
the density $\rho_{U_p}(\alpha)$ would not change. However, since the
eigenvalues of $D$ would be related to $\alpha$ differently, we would have
$\rho(\lambda=0)\neq 0$ for any $G$. The chiral transition would disappear. 
A similar effect can be seen by changing the sign of the gauge action.
Of course we have used the natural sign of the Wilson gauge action that is 
necessary for the continuum limit to emerge from a lattice action. Thus 
the staggered fermion phases and the natural sign of the Wilson action in 
eq.(\ref{eq:plaqact}) have conspired together to produce the chiral 
transition.

  Finally, it is interesting to know if one could extend such a 
calculation to the more interesting large $N$ phase transitions of Douglas
and Kazakov\cite{Douglas}, which unlike the transition studied here
is a continuum transition. Since the topology of space-time plays an 
interesting role in this continuum transition, including fermionic variables 
becomes non-trivial.

\section{Acknowledgment}

  I would like to thank Andrei Matytsin for helpful conversations. I also
would like to thank Norman Christ, Suzhou Huang and Uwe Wiese for suggestions 
and comments.

\begin{figure}[b]
\hskip0.5in
\epsfxsize=130mm
\epsffile{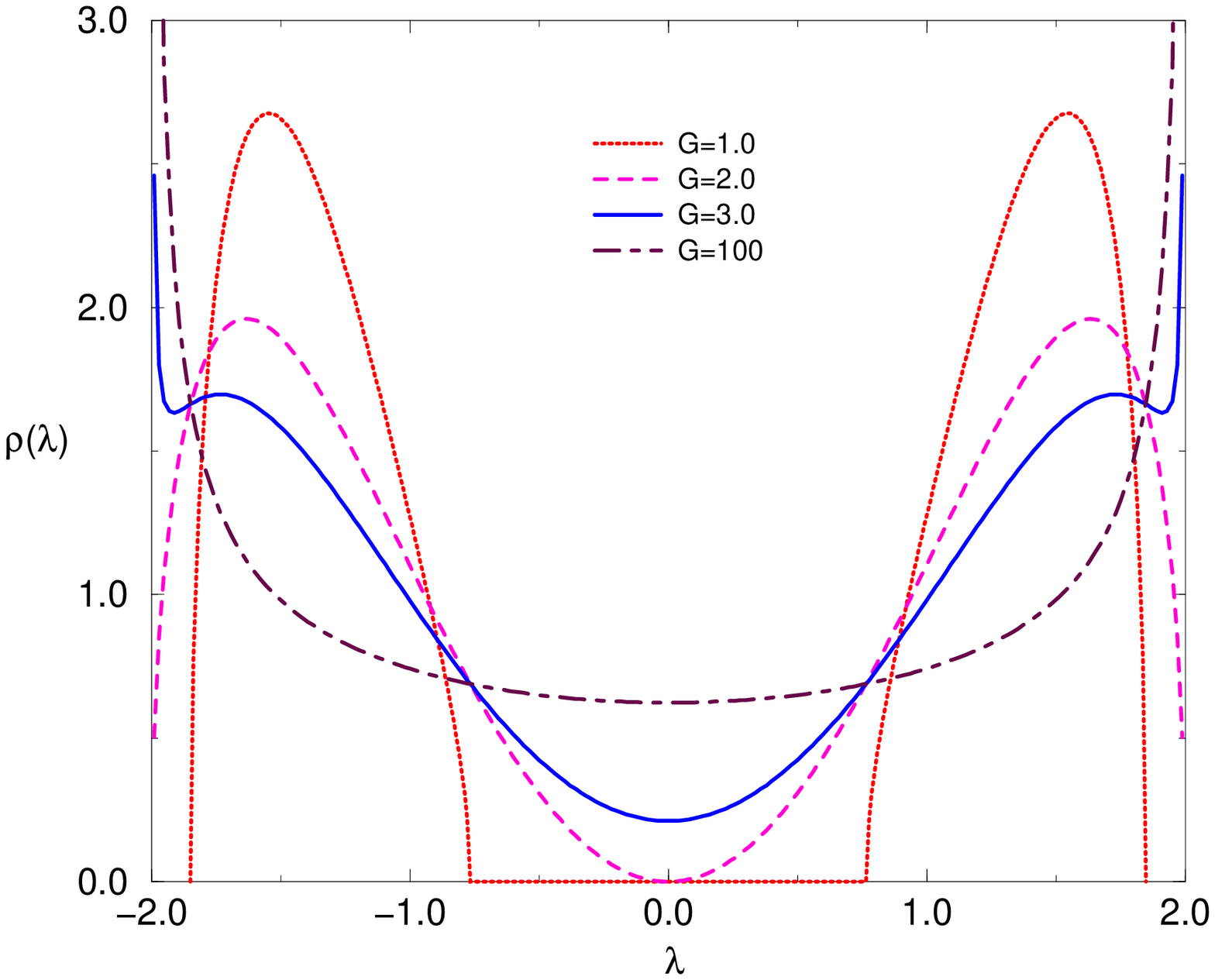}
\caption{
This plot of the eigenvalue density of $D$ defined in eq.(17)
for various values of the coupling $G$. At finite N, $D$ has $4N$ eigenvalues.
In the large $N$ limit these can be naturally be described by the density 
given above which is normalized to $4$. $G=2$ is the critical coupling.}
\end{figure}

\end{document}